\documentclass[conference]{IEEEtran}

\usepackage{tabularx}
\usepackage{algorithm}
\usepackage{algorithmic}
\usepackage{enumitem}
\usepackage{cleveref}

\usepackage{graphicx}
\usepackage{amsmath}
\ifCLASSINFOpdf
\else
\fi
\begin{document}
%
\title{Intelligent Perioperative System: Towards Real-time Big Data Analytics in Surgery Risk Assessment}

\author{\IEEEauthorblockN {Zheng Feng, Rajendra Rana Bhat, Xiaoyong Yuan, Daniel Freeman, Tezcan Baslanti, Azra Bihorac, Xiaolin Li\textsuperscript}
\IEEEauthorblockA{National Science Foundation Center for Big Learning\\
University of Florida, Gainesville, Florida 32603--0250\\
\{fengzheng, rbhat, chbrian, freemandaniel\}@ufl.edu, tezcan@phhp.ufl.edu, abihorac@anest.ufl.edu, andyli@ece.ufl.edu}
}


%


 \maketitle

\begin{abstract}
Surgery risk assessment is an effective tool for physicians to manage the treatment of patients, but most current research projects fall short in providing a comprehensive platform to evaluate the patients' surgery risk in terms of different complications. The recent evolution of big data analysis techniques makes it possible to develop a real-time platform to dynamically analyze the surgery risk from large-scale patients information. In this paper, we propose the Intelligent Perioperative System (IPS), a real-time system that assesses the risk of postoperative complications (PC) and dynamically interacts with physicians to improve the predictive results. In order to process large volume patients data in real-time, we design the system by integrating several big data computing and storage frameworks with the high through-output streaming data processing components. We also implement a system prototype along with the visualization results to show the feasibility of system design.

\end{abstract}

\vspace{2mm}

\begin{IEEEkeywords}
Perioprative risk prediction, Real-time processing, Big data analysis, Precision medicine
\end{IEEEkeywords}

%
\IEEEpeerreviewmaketitle

\section{Introduction}

According to the study, an average American is expected to undergo seven surgeries in his lifetime. Each year in US, at least 150,000 patients die and 1.5 million develop certain forms of medical complications within 30 days after surgeries~\cite{weiser2008estimation}~\cite{lee2008number}. It could potentially save thousands of lives by just reducing the postoperative complications (PC) by 20\% ~\cite{hall2009does}. Postoperative complications often lead to higher healthcare cost, adverse long-term stress, and other health issues. Among them, the Sepsis (SEP) and Acute Kidney Injury(AKI) are some complications that cause significant long-term morbidity and mortality~\cite{thottakkara2016application}. However, the mortality of SEP and AKI can be lessened by various preventive therapies based on physicians' risk assessment.

In prior research, the surgery risk scores are often subjectively assessed by physicians, and hence may suffer from inaccuracy. The presence of the large volume information-rich electronic health records (EHRs) also overwhelms physicians to comprehend every detail of a patient's profile. In addition, the characteristic of EHR data including high dimensionality, sparsity, and heterogeneity, makes it difficult to utilize them for modeling the perioperative risk, especially when they are applied in traditional statistical models. The viable alternative is utilizing data-friendly machine learning models that built on top of various features derived from data engineering approaches~\cite{collins2014tripod}. By applying these data on distributed streaming data processing framework, the real-time perioperative risk prediction is able to perform after aggregating and transforming the EHR data from different data sources. These techniques along with physicians' domain knowledge facilitate existing clinical decision support systems and improve patient-centered outcomes.

The integration of the analytic models and big-data techniques is a challenge in real clinical practice, owing to the complexity of processing real-time streaming data. The accuracy of the predictive models depend on domain expertise for feature selection process. Furthermore, traditional feature engineering approaches often scale poorly when facing the large volume EHR data from different sources, therefore missing the opportunity to discover novel patterns in data.

In this paper, we develop a real-time perioperative complication (PC) risk assessment system by using streaming EHR data. Furthermore, it calculates the risk scores for each new patient with high accuracy. IPS facilitates doctors to develop preventive strategies depending on the timely and accurate identification of the greatest perioperative complication risks for patients. It builds on open-source frameworks and runs various statistical and machine learning prediction models to provide accurate, automatic, and personalized perioperative risk assessment. 

\begin{figure*}[!t]
\centering
  \includegraphics[width=0.75\linewidth]{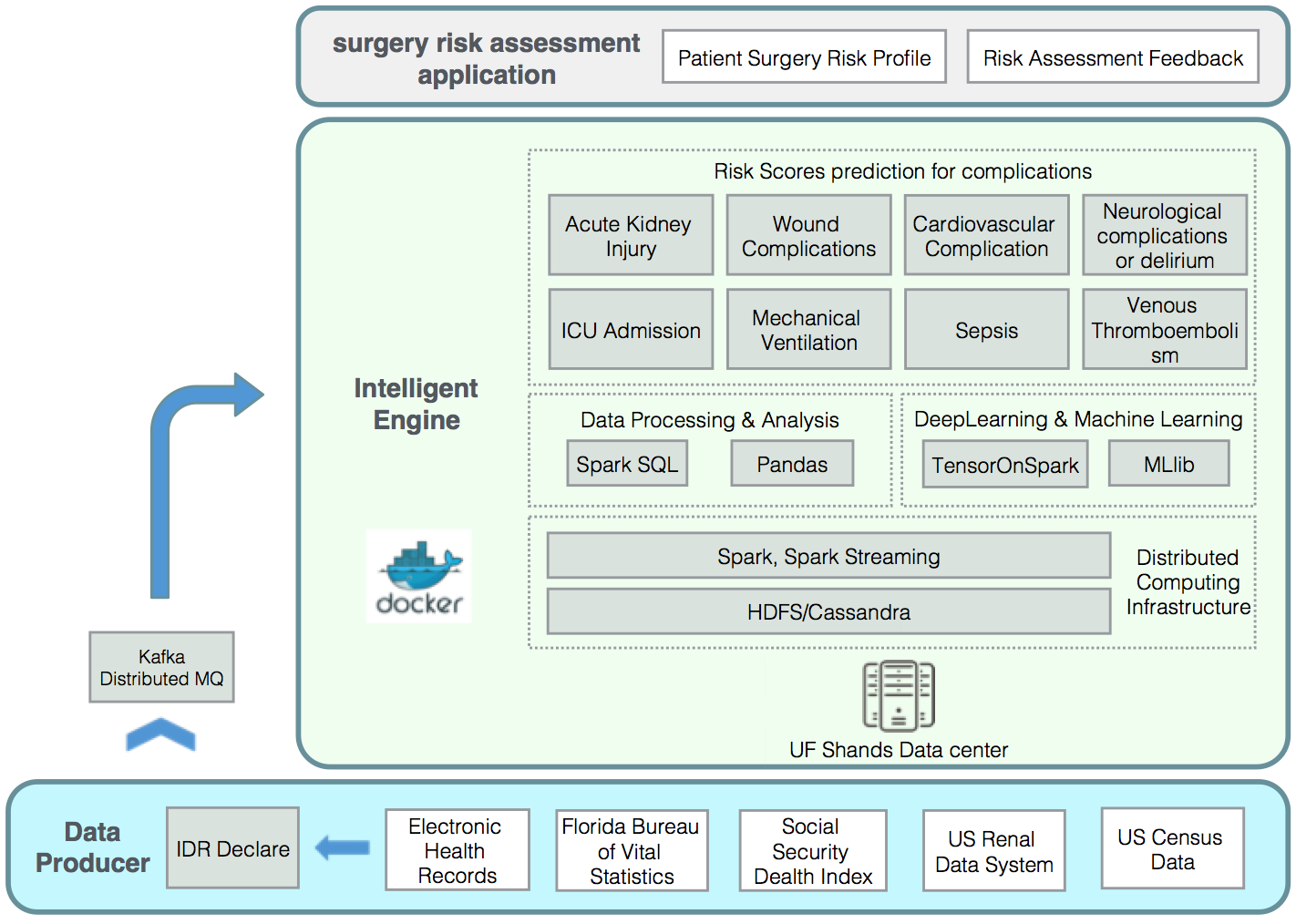}
  \caption{The Architecture of IPS. The system is designed into three components, including Data Provider, Intelligent Engine, and Application Client. The communication among different components is implemented by using distributed message queue and remote database access. All the components in Intelligent Engine are designed to work on separated Docker containers. The distributed Intelligent Engine cluster is deployed on UF Shands Data Center.}
  \label{fig:architecture}
\end{figure*}

\section{Related Work}
In this section, we review the existing research works of predicting and identifying complications or chronic diseases that applying statistical and machine learning models on EHR data. EHR data usually comprise of biological, anatomical and physiological data. They can be unstructured or semi-structured and come from various different sources such as patients' demographic information, discharge information, diagnostic/vital sign notes and check-in/check-out information.

For phenotyping identification, algorithmically automated EHR-based phenotyping by using Inductive logic programming (ILP) has been developed~\cite{peissig2014relational}. This relational machine learning (ML) model provider prediction for nine phenotypes, and it gives better results (in AUROC) compare to other non-relational approaches such as PART (p = 0.039), J48 (p = 0.003), and JRIP (p = 0.003). Similar experiments of predicting heart failure rate model using EHR data were conducted by M. Panahiazar, et al.~\cite{panahiazar2015using}. The new model comprising the Random Forest (RF) and Logistic Regression (LR) is evaluated on the benchmark of the standard Seattle Heart Failure Model. This model is applied to Mayo's Clinic data sets and it performs with better accuracy (11\% increase in AUC) and better prognostic prediction performance (8\% improvement in AUC), compared to the existing models after incorporating 26 more co-morbidities. In a different study, cardiovascular autonomic neuropathy detection was developed for diabetes patients using the ensemble methods including AdaBoost and Bagging(based on J48)~\cite{kelarev2012empirical}. This model is applied to the datasets collected from Diabetes Complications Screening Research Initiative at Charles Sturt University for the detection of Cardiovascular Autonomic Neuropathy. 

Recently, some applications are developed to extract novel probabilistic interdependence among disease-associated risk factors in various epidemics by applying the probabilistic framework on EHR data. The probabilistic frameworks EpiDefend and EpiAttack are proposed to identify and target the flu outbreak~\cite{dawson2015detecting}. This probabilistic model comprises of Dynamic Markovian Bayesian Network, Particle Filter, and text mining algorithms. This model applies on time series data such as WSARE datasets. Text mining algorithms are applied to determine anthrax epidemic by screening through the telephonic keyword. A similar approach of predicting pancreatic cancer is applied on different datasets including PubMed knowledge and EHR records by constructing a weighted Bayesian Network Inference (BNI) model~\cite{zhao2011combining}. In this study, twenty common risk factors were extracted from Pubmed knowledge to develop the BNI model called iDiagnosis for predicting pancreatic cancer. Compared to other machine learning methods, iDiagnosis outperforms existing machine learning models including k-Nearest Neighbor and Support Vector Machine. The probabilistic framework is also utilized in building real-time predictors for mortality and readmission~\cite{cai2015real}. This framework is built on Bayesian Network and applies to laboratory and administrative data of the patients, including 32,634 patient's records from the emergency department of Sydney metropolitan hospital in the span of 3 years. The average accuracy and AUC of the model are 0.80 and 0.82, respectively. With this model, they draw the trajectory of the patients and subsequently get some inference results including expected discharge, death, and readmission. 

\section{The Intelligent Perioprative System Framework}
This paper proposes a real-time intelligent perioperative system that periodically collects the EHR data of patients, and performs data integration, variable generation, surgical risk scores prediction, and risk scores visualization. It supports health professionals for their treatment evaluation and decision making. To fulfill the security requirements of University of Florida Health Integrated Data Repository (IDR) and build the system with high flexibility and scalability, we designed our system into separate components based on different roles of IDR and physicians. Each component is deployed on the separated server or platform and works in different trust region. The communication among different components is protected by multiple encrypting and security schemes. To process the data in real-time, the major data processing logic and prediction model are built on distributed subcomponents, and each subcomponent works individually with highly efficiency.  As shown in Fig~\ref{fig:architecture}, we designed our system into three components, including Data Provider, Intelligent Engine, and Application Clients. The Data Provider is a component integrating several data sources from IDR and formulating all the data into patients' admission based data stream. After collecting and aggregating the data from IDR, the transformed patients' data stream is sent to the Intelligent Engine through distributed message queue. The Intelligent Engine periodically fetches the data from the message queue, processes the data by streaming working logic, and finally stores the results into NoSQL database for further interpretation and visualization. The last component builds on top of the intelligent engine is the Surgery Risk Assessment Application clients. The major functionalities of the application clients include patient surgery risk profiling and physician feedback on risk assessment. 

\begin{figure}[htb]  
  \includegraphics[width=\linewidth]{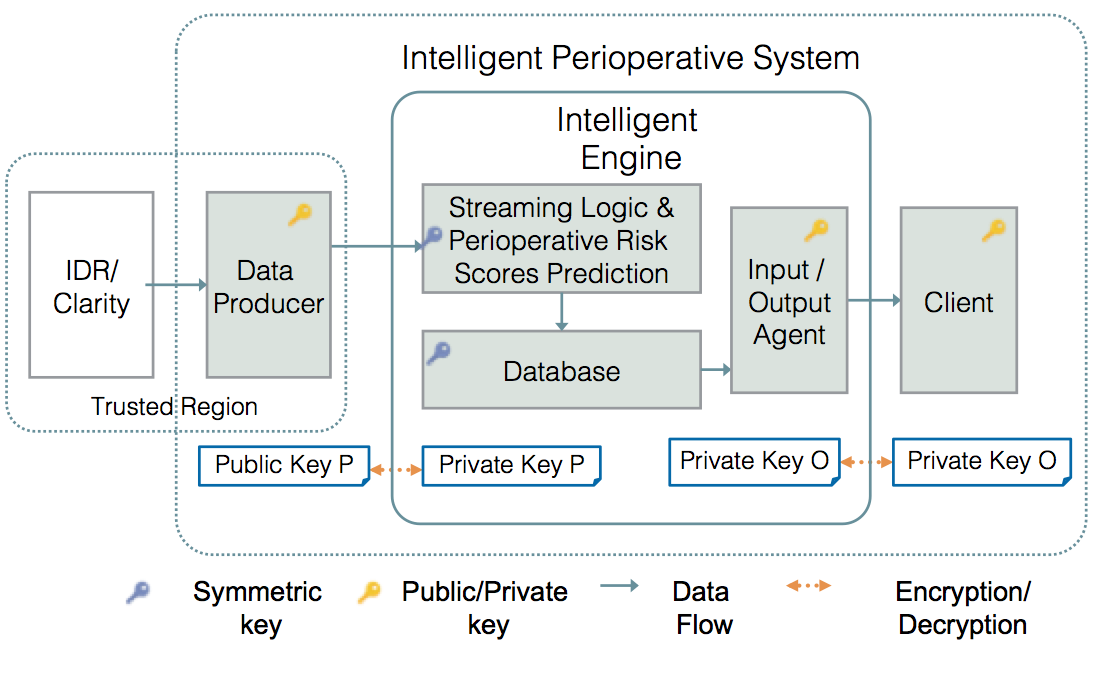}  
\caption{Security Design of IPS. Here we use three different schemes to ensure the system security while exchanging the sensitive data, including non-encrypted data exchanging for sub-components in same trusted region, public-key encryption applied among different subcomponents out of trusted region, and symmetric key encryption is applied in the communication between streaming logic and database. }
\label{fig:ips_security}
\end{figure}

\subsection{Security Schemes Design}
Because the processed data in the system contains sensitive health information of patients, the information security is the essential part of system design required by IDR. To ensure the system security, different schemes are proposed while the sensitive data is exchanged among different system components. The schematic diagram of the security design is shown in Fig~\ref{fig:ips_security}. First, the IDR maintains all the data from different sources and provides the data to the system whenever new records get in. To work with IDR, the data producer continuously checks on the directory of IDR for new data. During this process, the encryption of the message is not required because the Data Producer works in the same secure domain with IDR. But when data exchanges among servers that work in the different region, the communication is required to be secured. This happens when Data Producer sends data to the Intelligent Engine and the Application Clients access the database. In these situations, the system applies the public key infrastructure (PKI) to protect the exchange of the sensitive information. Data exchanging is encrypted through the Secure Sockets Layer (SSL) protocol and only the intended receiver can decipher the data by using the private key it possesses. The Streaming work logic and prediction models in Intelligent Engine can be deployed at different cluster from the database at some working environment, so the encrypted communication between these two subcomponents is required as well. Considering the mass data exchanging of them, we use the symmetric key encryption to encrypt the data, because compare with the public-key encryption algorithms, the symmetric-key algorithms are more efficient, and the symmetric key can be designated during the configuration before running. We also apply RestAPI~\cite{masse2011rest} for clients to communicate with our system. This avoids clients to directly talk with the system and may provide the opportunity for SQL injection and other malicious attacks. 

\begin{figure}[htb]  
  \includegraphics[width=\linewidth]{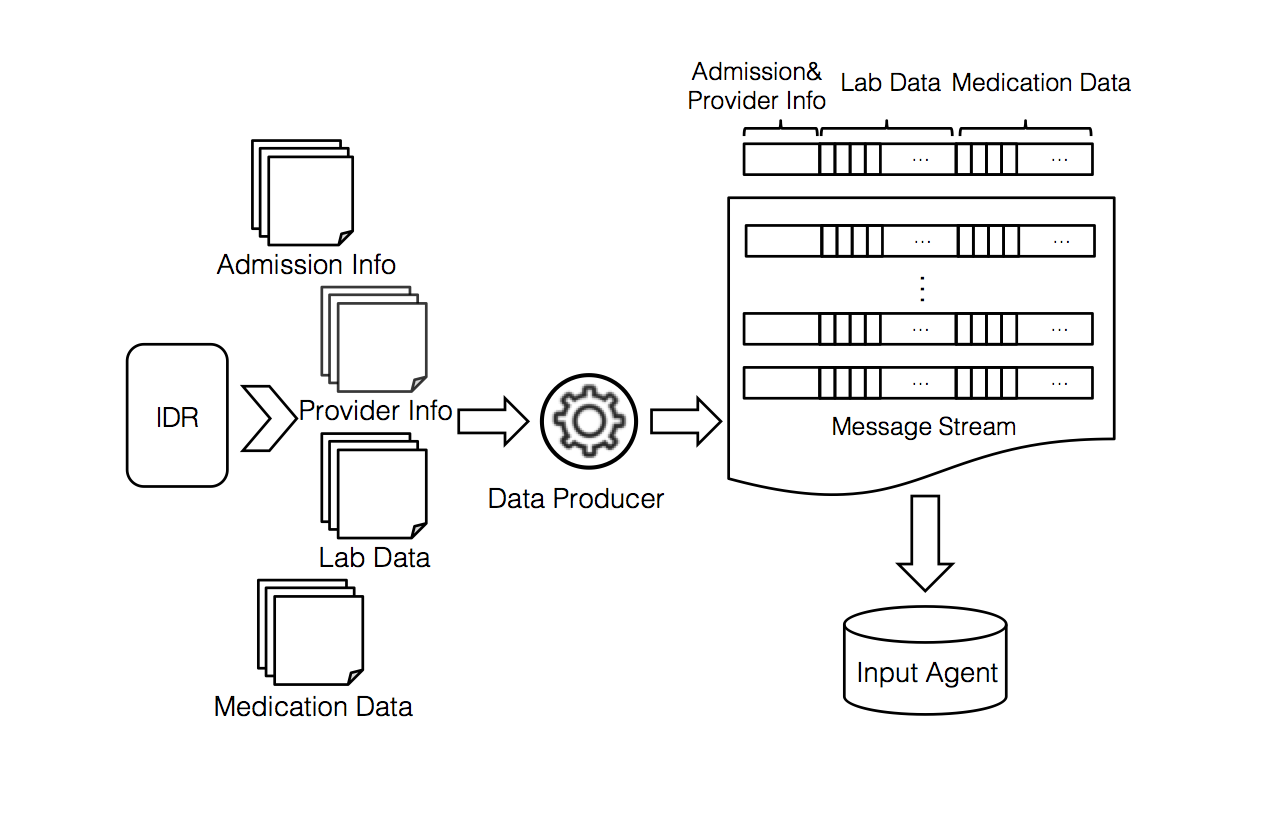}  
\caption{The Workflow of Data Producer. For each patient admission, there is at least 1 provider record, zero or multiple lab tests, and medication records. The Data Producer will collect and connect all the records and transform them into patient admission based records stream.}
\label{fig:ips_integration}
\end{figure}

\subsection{Data Producer}
The primary purpose of the Data Producer is gathering all the data files from different sources provided by IDR. After aggregating and connecting these data, it transforms them into patient based records stream. To this end, the Data Provider performs the process shown in Fig~\ref{fig:ips_integration}. All the data files provided by IDR come from the patients' Electronic Health Records, Florida Bureau of Vital Statistics, Social Security Death Index, US Renal Data System, and US Census Data. The major information used in the current surgery risk prediction models has four categories, including patient admission information, provider information, lab tests data, and medication data. For each patient admission, there is zero or multiple lab tests and medication records and at least 1 provider record. The Data Producer progress is scheduled to check the directories of IDR repeatedly in a configured time interval. At each interval, the Data Producer collects a new batch of data files, then connect all the records in different files based on the patient admission identity. All the records are transformed into JSON string and remotely wrote into the input agent of Intelligent Engine. The input agent of Intelligent Engine is implemented with Kafka~\cite{ranjan2014streaming} distributed message queue, so the producer is able to simultaneously write to multiple servers in Intelligent Engine cluster. 

\subsection{Intelligent Engine}
The Intelligent Engine consists of tools and modules for processing and analyzing the data stored in the NoSQL database or from streaming data source. Here we utilize the Spark~\cite{zaharia2016apache} as our distributed computational infrastructure, Cassandra~\cite{chebotko2015big} as NoSQL database, and HDFS~\cite{borthakur2007hadoop} as distributed file system solution. In order to tighten the system security, and make the deployment of the Intelligent Engine with the minimum configuration at the different production environment, we use Docker~\cite{merkel2014docker} container to host all the subcomponents of Intelligent Engine. 

The Intelligent Engine provides three functionalities shown in Fig~\ref{fig:architecture} including real-time complications risk scores prediction, batched model training with distributed machine learning/deep learning tools, and SQL based data analysis. First, the real-time risk scores prediction builds on the spark streaming infrastructure. The surgery risk prediction is the major functionality of the Intelligent Engine, the customed real-time prediction job constantly works on top of the spark streaming and periodically pulls the patient admission based JSON records from Kafka distributed message queue. The acquired patient records stream flows through the defined streaming work logic including the data engineering/preprocess, surgery risk prediction, and storage of the final results. All these three subcomponents in the streaming logic work individually with their own functional modules and we use the streaming interfaces to couple them together to work in a pipeline. The data engineering/preprocessing subcomponent first convert each patient JSON record into raw features. Then it transforms and remodels these raw features based on several predefined dictionaries to fit the input of 8 complications risk prediction models. The 8 
complications prediction models are currently designed to work independently from each other. But the multitask learning that share the common learning features are also supported. This gives the system opportunities for exploring the interrelationship among different complications. The batch model training is established on the distributed machine learning tools of Spark, including the Mlib for general machine learning tasks, and TensorOnSpark~\cite{smith2016scaling} for deep learning tasks. The general data analysis and processing tasks are performed by the SparkSQL. It executes on the powerful Spark distributed computational engine for the computational ability and NoSQL database for large amount unstructured data storage.

\begin{figure}[!ht]  
\centering
\includegraphics[width=\linewidth]{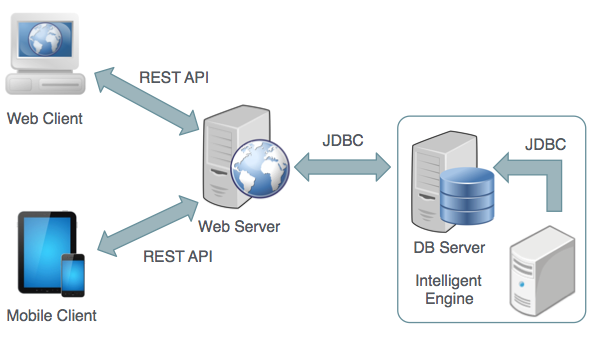}  
\caption{Client Interface Architecture. In the system, each client interface with server via RestAPI. RestAPI is deployed in Apache Tomcat web server.}
\label{fig:ips_application}
\end{figure}

\subsection{Application Clients}
The application component provides the real time display of patients' complete information. It includes the prediction of the likelihood of PCs as well as some descriptive information of patients for physicians. This component consists of web service, presentation, and visualization. In Web Service, we use RestAPI as the intermediate agent for the clients and Intelligent Engine. Various clients interact with RestAPI instead of directly talk to Intelligent Engine. Fig~\ref{fig:ips_application} provides the architectural design of the application clients components. For visualization, the module uses JavaScript based D3 visualization framework. It renders pie-chart graph that allowing the physician to increase or decrease the pie based on their evaluation of the prediction.  



\begin{figure}[!ht]  
\centering
\includegraphics[width=\linewidth]{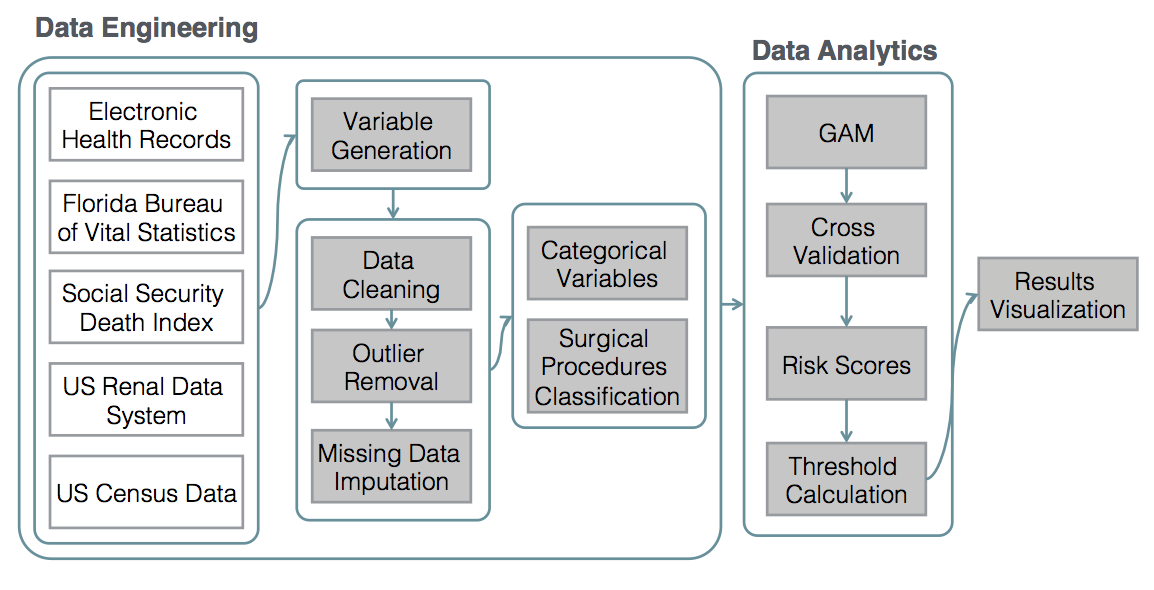}  
\caption{Implementation of IPS Prototype. The work logic of the implementation can be summarized in two parts, including the data preprocessing and data analytics.}
\label{fig:ips_analytic}
\end{figure}

\section{Intelligent Perioperative System Prototype Implementation}
\begin{figure}[!ht]  
\centering
\includegraphics[width=\linewidth]{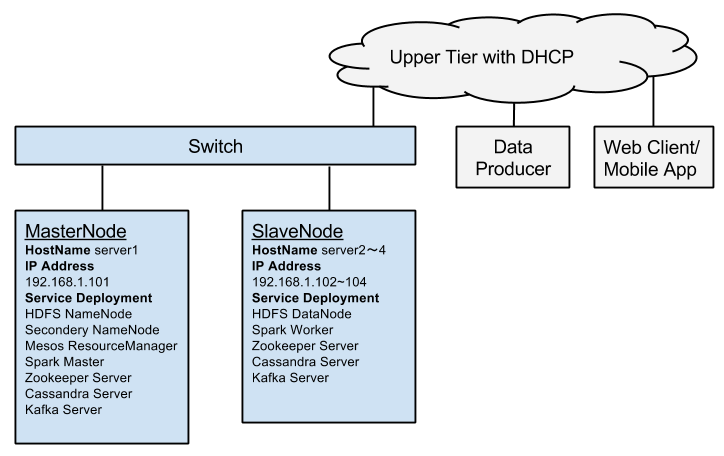}  
\caption{Configuration of IPS prototype. The cluster of Intelligent Engine is deployed on 4 servers and one of them serves as master node and the other 3 serve as slave nodes. The nodes in the cluster exchange the data through a gigabit switch. The DataProducer and clients communicate to the Intelligent Engine through Internet.}
\label{fig:ips_configuration}
\end{figure}

In the first stage of our work, we implement a prototype of the intelligent perioperative system to validate the feasibility of the proposed system design. In this section, we describe the major implementation of the system, as shown in Fig~\ref{fig:ips_analytic} the work logic of the implementation can be summarized into two parts, including the data engineering and data analytics. The data engineering contains all the work of data transformation and feature extraction. The data analytics describes the model training and the real-time prediction. At the end, we present the visualized results of a patient that shows his predicted surgery risk. When building the system, we use the configuration shown in Fig\ref{fig:ips_configuration}. we deployed the system on 4 x86 servers with CPU of E5-2695 X8, 256 GB memory, and 3 TB storage. One of them serves as the master node and other 3 serve as the slave nodes. The software environment is described in Table \ref{tab:ips_env}. With this setting, the system can reach the through-output of 5000 records per minute. And the average system delay for a single record is 60 s.

\begin{table}[htb]
\renewcommand{\arraystretch}{1.1}
\caption{\textbf{The software environment of IPS prototype}}
\label{tab:ips_env}
\centering
\begin{tabular}{l c c}
\hline
\textbf{Software} & \textbf{Version} \\
\hline
\textbf{JDK} & 1.8.0 &\\
\textbf{Python} & 2.7.12 &\\
\textbf{Scalar} & 2.10.1 &\\
\textbf{HDFS} & 2.6.4 &\\
\textbf{Spark} & 1.6.2 &\\
\textbf{Cassandra} & 3.9 &\\
\textbf{Kafka} & 2.10-0.10.0.1 &\\
\textbf{ZooKeeper} & 3.4.9 &\\
\hline
\end{tabular}
\end{table}

\subsection{Data Engineering}
All the data in the prototype is collected from University of Florida Health Integrated Data Repository, and it includes a single cohort of the patients admitted to Shands Hospital at the University of Florida from January 2000 and November 2010. Patients that aged 18 years and above, with the hospital stay of greater than 24 hours, are selected, giving the total of 50,314 patients. The patients having end-stage renal disease on admission and missing serum creatinine were excluded. In order to analyze the data, the raw data is preprocessed by a set of data engineering techniques to generate applicable features for the predictive model. The data engineering process includes variable generation, data cleaning, outlier removal, and missing data imputation. The variable generation extracts useful patients information (demographics, socioeconomics, operative information, and comorbidity related information) from raw data to generate a set of variables. 
Once data is collected, data cleaning including removal of outliers and imputation of missing data is executed. Mean imputation is applied for continuous variable whereas the missing category is created for nominal variables.  

\subsection{Data Analytics}
In current stage, the main target of the Data Analytics subcomponent is to apply the surgery risk predictive models on processed data stream. The complete work flow is shown in Fig~\ref{fig:ips_analytic}. The predictive model applied in Data Analytic is Generative Additive Model (GAM) model~\cite{thottakkara2016application}. All the GAM models of 8 complications are pre-trained and encapsulated in R packages. For each complication, the model produces the predicted surgery risk score along with the important contributing risk factors.
The output risk scores categorize patients into low-risk and high-risk groups by employing a cutoff (the threshold value) evaluated by maximizing the Youden index. The calculated cutoff values for all 8 complications are AKI (0.35), ICU (0.35), MV (0.13), WND (0.10), CV (0.07), NEU (0.07), SEP (0.06), VTE (0.03), respectively. To ensure that we are able to select the relative better model, five-fold cross-validation is employed and the corresponding performance metrics are reported. 

\begin{figure}[!ht]  
\centering
\includegraphics[scale=0.30]{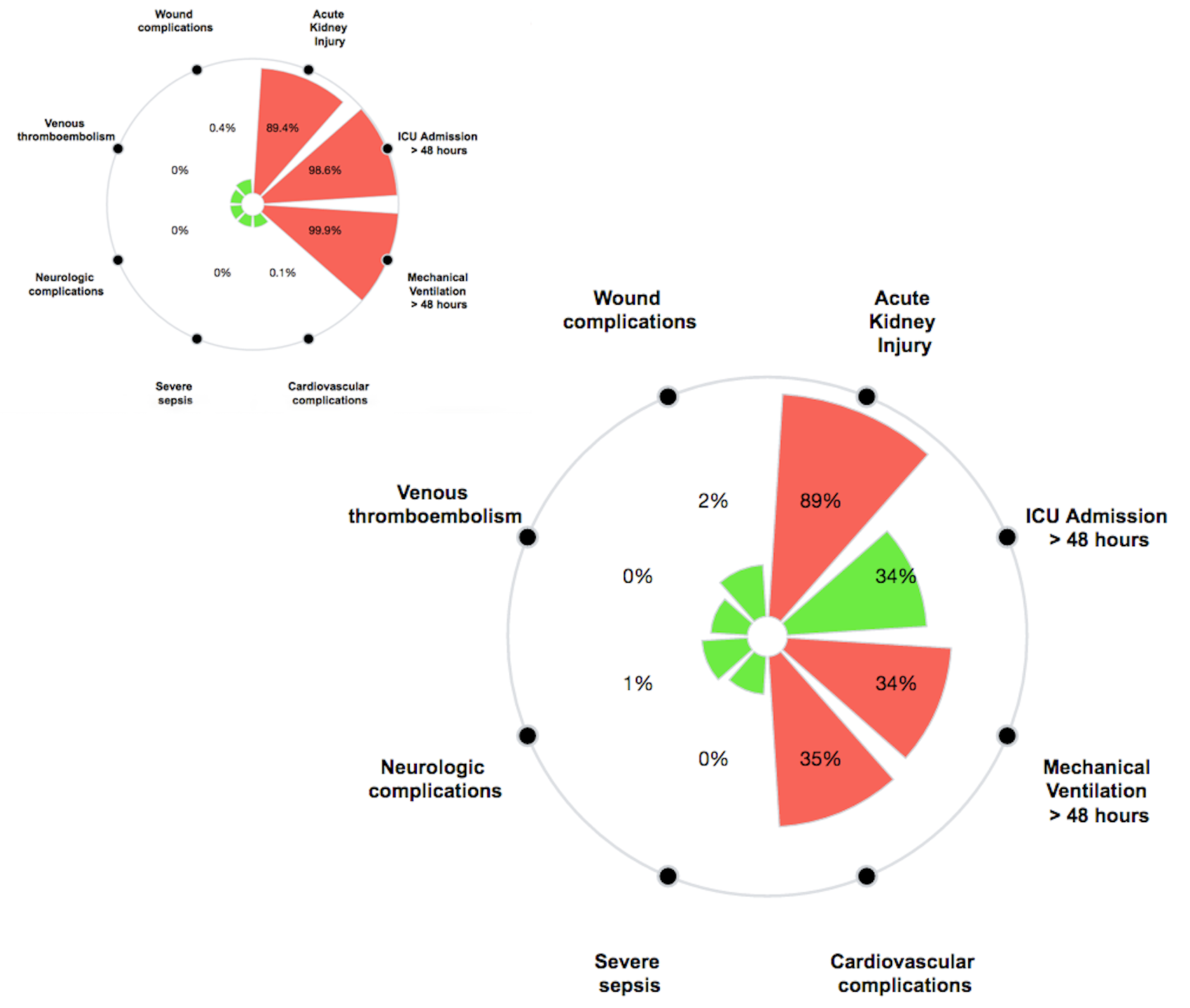}  
\caption{Web Client for IPS. Computer generated risk assessment with final patient's risk assessment is displayed.}
\label{fig:ips_web}
\end{figure}

\subsection{Results and Visualization}
IPS implements two clients for physicians to conveniently access and interact with the system in any moment. It includes Mobile app client and thin web client, and they both exchange the information via RestAPI. The RestAPI provides the interface for various clients and avoids the clients to query the data from the system core directly. This also augments the scalability of the system because it decouples the system from different clients of various platforms.

\subsubsection{Web Client}
The web client provides physicians a series of services to facilitate them monitoring the immediate surgery risks of their patients. This includes the Email notification for the new status of patients, the patients' profile sketch generation, and visualization of each predicted surgery risk scores with pie chart. The example of predicted scores visualized a pie-chart is shown in Fig~\ref{fig:ips_web}.  In this pie-chart, the IPS system predict the risk scores of a patient for postoperative complications includes Acute kidney injury, Cardiovascular complications, Intensive care unit admission $>$ 48 hours, Mechanical ventilation $>$ 48 hours, Neurologic complications, Sepsis, Venous thromboembolism, and Wound complications. 

\begin{figure}[!ht]  
\centering
\includegraphics[scale=0.50]{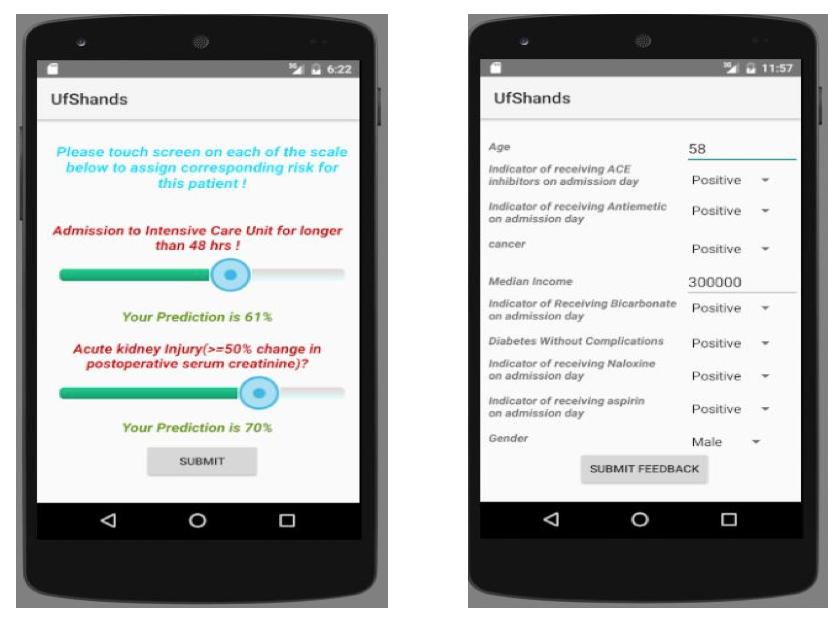}  
\caption{Mobile Client for IPS. AKI and ICU scores are displayed in left screen and physician registration page in right screen.}
\label{fig:ips_mob}
\end{figure}

\subsubsection{Mobile App}
The mobile client implements on Android Operating System. Compare to the Web client, the mobile client contains the functionality of pushing notification through Google Cloud Messaging (GCM) to the physicians once the results of their patients are available. Fig~\ref{fig:ips_mob} shows few screen shots of our mobile client.

\section{Conclusion And Future Works}
In this paper, we developed IPS, a distributed real-time system for perioperative risk prediction. The system applies the predictive risk models for major complications by using EHR data. We implemented a prototype to validate the system design, and we believe it is the first real-time perioperative risk prediction system for the clinical usage. Motivated by this idea, we are optimistic that this streaming analytics paradigm shall be an effective tool for the clinics and hospitals surgery management in the US.
In next stage of our work, we plan to replace the current individually trained complication models with an ensembled multi-task model and integrate it into the current streaming system. This takes into account the inter-relationship of all the complications and gives better generalizing ability for each complication by sharing the common features of all the complications.

\section*{Acknowledgement}
The work presented in this paper is supported in part by National Science Foundation (grants CNS-1624782, OAC-1229576, CCF-1128805) and National Institutes of Health (grant R01-GM110240).

\bibliographystyle{IEEEtran} 

\bibliography{references}

\end{document}